# In-pixel automatic threshold calibration for the CMS Endcap Timing Layer readout chip

H. Sun,[a,b,1] D. Gong,[b] C. Edwards,[c] G. Huang,[a] X. Huang,[a,b,1] C. Liu,[b] T. Liu,[b] T. Liu,[c] J. Olsen,[c] Q. Sun,[c,2] J. Wu,[c] J. Ye,[b] L. Zhang,[a,b,1] W. Zhang,[a,b,1]

[a] *Central China Normal University,*
   *Wuhan, Hubei 430079, PR China*

[b] *Southern Methodist University,*
   *Dallas, TX 75275, USA.*

[c] *Fermi National Accelerator Laboratory,*
   *Batavia, IL 60510, USA.*

   E-mail: qsun@fnal.gov

ABSTRACT: We present the implementation and verification of an in-pixel automatic threshold calibration circuit for the CMS Endcap Timing Layer (ETL) in the High-Luminosity LHC upgrade. The discriminator threshold of the ETL readout chip (ETROC) needs to be calibrated regularly to mitigate the circuit baseline change. Traditional methods need a lot of communication through a slow control system hence are time-consuming. This paper describes an in-pixel automatic scheme with improvements in operating time and usability. In this scheme, a sample-accumulation circuit is used to measure the average discriminator output. A binary successive approximation and linear combination scan are applied to find the equivalent baseline. The actual calibration procedure has been first implemented in FPGA firmware and tested with the ETROC front-end prototype chip (ETROC0). The calibration circuit has been implemented with Triple Modular Redundancy (TMR) and verified with Single Event Effects (SEEs) simulation. A complete calibration process lasts 35 ms with a 40 MHz clock. In the worst case, the dynamic and static power consumption are estimated to be 300 μW and 10.4 μW, respectively. The circuit design, implemented in a 65 CMOS technology, will be integrated into ETROC2, the next iteration of the ETROC with a 16×16 pixel matrix.

---

[1] Visiting scholars at SMU and performed this work at SMU.
[2] Corresponding author.

# Contents

## 1. Introduction

The MIP (minimum ionizing particle) Timing Detector (MTD) is a new timing detector being developed for the Compact Muon Solenoid (CMS) experiment in the High-Luminosity Large Hadron Collider (HL-LHC) [1], with the mission of disentangling the approximately 200 nearly-simultaneous "pileup" interaction in each bunch crossing. It also enables new capabilities for charged hadron identification and the search for long-lived particles. As a sub-detector of MTD, the Endcap Timing Layer (ETL) will be instrumented on each side of the interaction region, with a two-disk system of Low Gain Avalanche Detector (LGAD) [2-4]. The LGAD provides two hits per track with 30 to 40 ps time resolution. The two-hit design relaxes the time resolution of each disk to 50 ps level, which is a challenge for the front-end electronics constrained by a low power budget.

The LGAD optimized for ETL has a 16 × 16 pixel matrix with each pixel 1.3 × 1.3 mm$^2$. The ETL Readout Chip (ETROC) being developed provides signal processing and data readout logic to match the LGAD. One ETROC has a 16×16 pixel matrix and global circuits at the bottom of the pixel matrix, as shown in Figure 1. Each pixel channel in ETROC consists of a preamplifier, a discriminator, a time-to-digital converter (TDC) [5], an in-pixel readout block, a local I$^2$C-based slow control block, and a digital-to-analog converter (DAC) [6] for discriminator threshold setting.

In the ETL, an incident charged particle deposits energy in LGAD and generates charge through ionization. LGAD amplifies the charge, and the induced current flows into the preamplifier in ETROC. The preamplifier converts the input current into a voltage pulse. The discriminator compares the voltage signal against a threshold generated by the 10-bit DAC. The TDC digitizes the time information of the discriminator output, including Time-Of-Arrival (TOA) and Time-Over-Threshold (TOT). TOT is an estimate of the input signal size and is used for the time walk correction. Each channel's detailed hit information (TOA and TOT) is written into the in-pixel readout circuit and read out once a Level-1 trigger is received [1]. A charge injector is also included to facilitate the test and calibration.

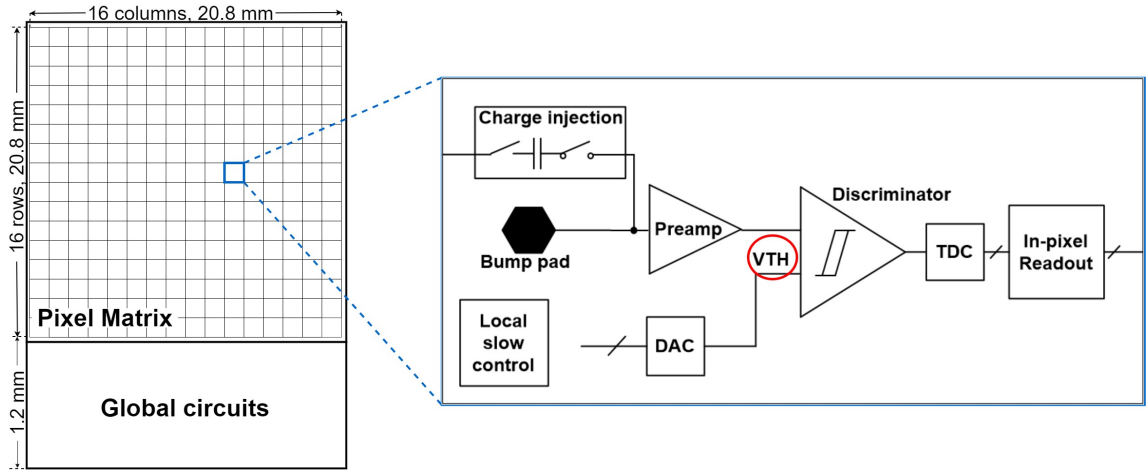

**Figure 1.** Block diagram of ETROC with the single pixel channel in detail.

Like other detectors used in High Energy Physics (HEP) experiments, a proper threshold is important for the ETROC discriminator to mitigate dispersion and obtain optimal performance. Traditional HEP detectors require dedicated off-chip threshold calibration [7-11]. The off-chip process is usually time-consuming, requiring minutes to hours to complete. We proposed an in-pixel automatic threshold calibration for ETROC, significantly simplifying the procedure and reducing the calibration time during detector operation.

In Section 2, the motivation of the in-pixel threshold calibration is briefly described. Section 3 discusses three possible approaches. The chosen approach and the verification in FPGA are depicted in Section 4. Section 5 presents the circuit implementation. A conclusion is drawn in the last section.

## 2. Motivation of the in-pixel threshold calibration

A proper discriminator threshold, usually a few millivolts above the equivalent baseline, is important for optimal performance. The equivalent baseline is defined as the applied threshold when the discriminator output falls in the transition region (between logic 0 and logic 1). It is dominated by the preamplifier baseline and impacted by the input-referred offset voltage of the discriminator. The impact of the threshold on the performance was studied with SPICE simulation, where an LGAD signal set from WeightField2 [12] simulation with a large range of amplitudes was applied to input ETROC analog front-end [13, 14]. The charge variation of the 1000 events in the signal set introduces a significant time walk because of Landau distribution, and TOT-based time walk correction was applied. Figure 2(a) shows the simulated time resolution as a function of the discriminator threshold; a higher threshold leads to a better time resolution because it suppresses the small input signals. However, a lower threshold allows a better efficiency, as shown in Figure 2(b). The optimal threshold lies at the middle range (7~9 mV) in this simulation.

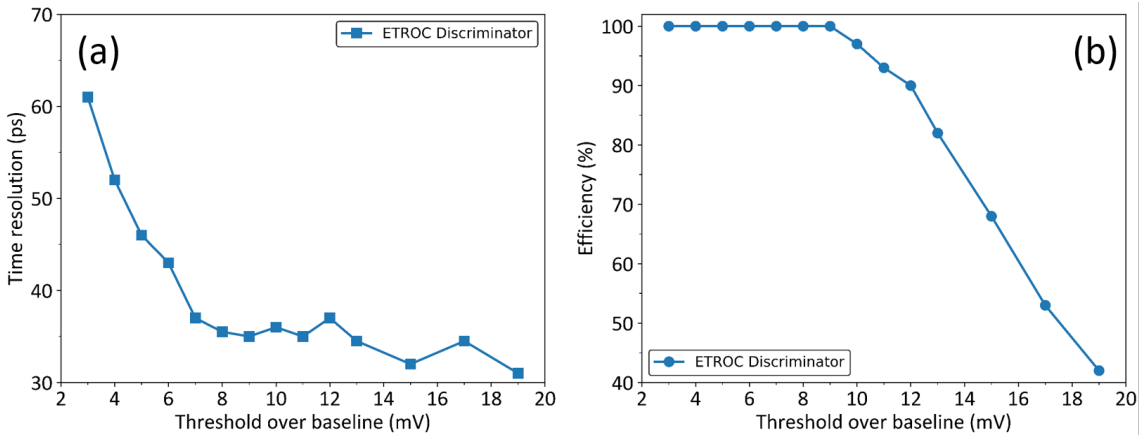

**Figure 2.** The dependence of discriminator threshold on time resolution (a) and efficiency (b).

As the dominator of the equivalent baseline, the preamplifier baseline can disperse and change over time due to many factors. The temperature coefficient of the baseline shift is about -1 mV/°C from the post-layout simulation. A higher preamplifier bias (power) setting leads to a higher baseline. In addition, the preamplifier baseline in different pixels on the same die or different dies disperses due to process variations and mismatch. From a Monte Carlo analysis where both global (process) variation and local (mismatch) variation are included, the standard deviation of the preamplifier baseline is about 27 mV, indicating that the threshold calibration should be performed pixel by pixel. Besides, the LGAD leakage current contributes to the preamplifier baseline, given that the LGAD connects to the preamplifier with DC-coupling.

A prototype of the ETROC series, named ETROC0, includes a single-channel analog front-end and has been extensively tested under different conditions [13, 14]. Those tests reveal that ETROC0 survives up to 1 MGy Total Ionization Dose (TID), and the TID effect does affect the equivalent baseline, as expected. For example, when ETROC0 chips were irradiated with x-ray by about 15 kGy, the equivalent baseline was observed to decrease by about 8.4 mV.

All the above factors indicate that the equivalent baseline disperses or changes due to process, setting, and operational environment. Therefore, the threshold calibration should be performed regularly to obtain the correct functionality and reasonable performance. Though an off-chip threshold calibration is traditionally feasible for the HEP experiment, an in-pixel threshold calibration significantly saves calibration time and effort. The proposed scheme aims to find the equivalent baseline and allows users to apply an optimal threshold easily.

## 3. Principles of the in-pixel approaches

This section presents three possible approaches for ETROC in-pixel threshold calibration. We also compare their features and explain how we make a choice.

### 3.1 S-curve measurement

S-curve measurement is a traditional calibration procedure extensively used by many HEP detectors [7, 9, 11]. In this procedure, a reference charge is injected into the preamplifier while the discriminator threshold is scanned around the preamplifier pulse peak. The discriminator output is constantly low when the threshold is higher than the pulse peak and constantly high when the threshold is below the preamplifier baseline. In both cases, the discriminator pulse counts are 0. The discriminator is fired by the injected charge with the threshold between the peak and the baseline. When the threshold is in the vicinity of the peak, the discriminator is fired peak and the baseline.

randomly by the noise. The average discriminator output is calculated and illustrated by the S-curve versus the threshold. The transition region on the S-curve represents the preamplifier pulse peak, and the analog front-end noise can be found through S-curve analysis.

In the S-curve measurement for ETROC threshold calibration, the discriminator threshold is described as charge (unit: fC) [15,16]. At least two runs of the S-curve measurements are needed to derive the preamplifier gain (unit: mV/fC) and set a threshold. Figure 3(a) illustrates the conceptual preamplifier signal with threshold scanning for two S-curve measurements. The measured S-curves when the threshold crosses the pulse peak are shown in Figure 3(b). Figure 3(c) illustrates the counting results versus an extended threshold scanning range and will be detailed in Section 3.4. The gain of the preamplifier could be calculated as follow:

$$Gain = \frac{TH_{upper} - TH_{lower}}{Q_{upper} - Q_{lower}}, \quad (3.1)$$

where $TH_{upper}$ and $TH_{lower}$ are the measured threshold for the charge $Q_{upper}$ and $Q_{lower}$, respectively, and $Q_{upper}$ is larger than $Q_{lower}$. The target threshold, $TH_{set}$, is given by:

$$TH_{set} = TH_{lower} - (Q_{lower} - Q_{set}) * Gain, \quad (3.2)$$

where $Q_{set}$ is the charge at which the users would like to set the threshold.

### 3.2 Noise rate scan

The second possible approach is noise rate scan. It allows finding the equivalent baseline through the discriminator pulses triggered by the noise without any input signal. Figure 4(a) illustrates the preamplifier output baseline with fluctuation due to noises. As for an arbitrary threshold, some noise pulses will always cross the threshold since the noise amplitude distribution is Gaussian; but the noise rate will vary with the threshold [17]. The noise rate reaches its maximum with the threshold level set to zero relative to the equivalent baseline. For ETROC in-pixel threshold calibration, the discriminator pulses can be counted in a time window to get the noise rate. A conceptual noise distribution is shown in Figure 4(b). The peak of the noise rate sits at the equivalent baseline, and the width of the Gaussian distribution represents the noise amplitude.

### 3.3 Sample accumulation

Another approach for the ETROC threshold calibration is sample accumulation. The basic idea of this approach is to measure the average discriminator output versus scanned thresholds and find the transition point without input signals injected. In each scan step, the discriminator is sampled at a given frequency, and then the samples are accumulated in a measurement window. The discriminator output is constantly high when the threshold is lower than the baseline and low when the threshold is above the baseline. The scanned threshold crosses the preamplifier output baseline, as shown in Figure 4(a), and the accumulated samples of the discriminator output versus threshold scanning are shown in Figure 4(c). The transition region represents the noise amplitude, and the equivalent baseline locates around half of the maximal accumulated numbers.

### 3.4 Comparison of approaches

Table 1 summarizes the three threshold calibration approaches discussed above. The S-curve measurement requires charge injection to generate signals and finds the preamplifier pulse peak. The threshold locating at the transition point is described as a reference charge. The threshold scan should be conducted downward by a user-defined step and usually stops at a middle value when the threshold is between the peak and the baseline of the preamplifier signal. The threshold scan range could be extended to cross the preamplifier baseline, as the green dash-dot curve

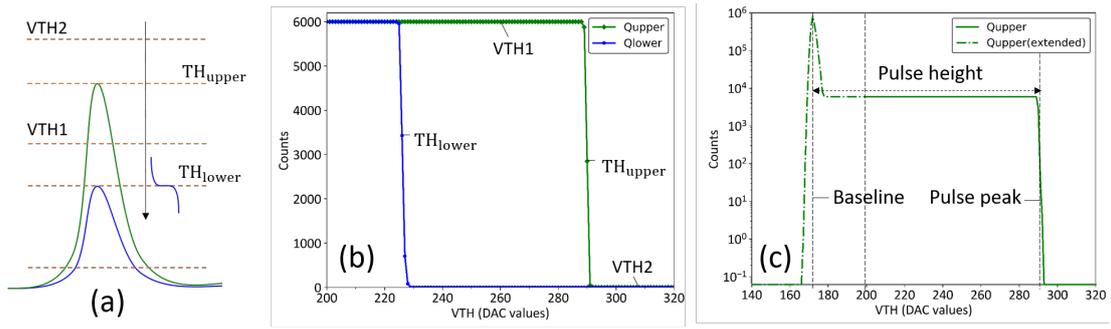

**Figure 3.** S-curve measurement: preamplifier signals with the threshold scanning (a), the discriminator pulse counts versus the typical threshold scanning for different input charges (b), and the discriminator pulse counts versus the extended threshold scanning range when crossing the preamplifier baseline (c).

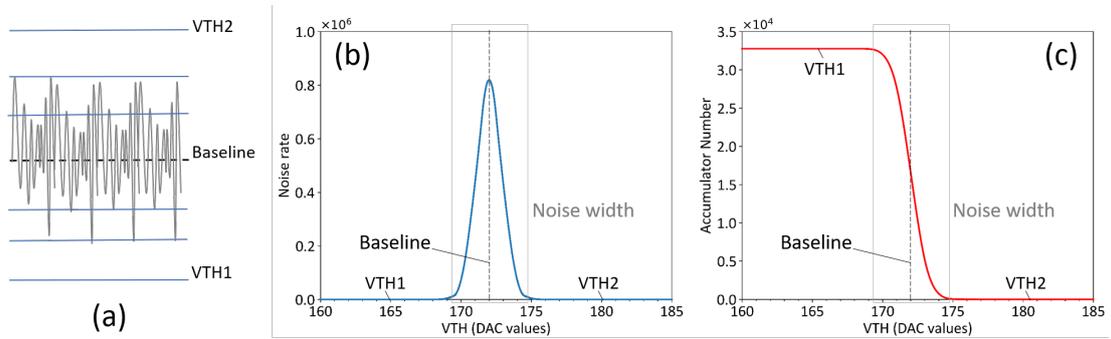

**Figure 4.** Preamplifier output baseline with fluctuating noise pulses (a), the noise rate versus the threshold scanning in the noise rate scan approach (b), and the accumulator number versus the threshold scanning in the sample accumulation approach (c).

Table 1. In-pixel threshold calibration approaches comparison

| Approaches | Pros. | Cons. |
|---|---|---|
| S-curve measurement | Threshold corresponding to a reference charge; Simultaneous pulse height measurement. | Requiring charge injection. |
| Noise rate scan | Obvious baseline locating at the noise rate peak. | Unpredictable maximal noise rate. Requiring linear scan and being time-consuming. |
| Sample accumulation | Fast scan with binary SAR algorithm. | Requiring extra care due to the asynchronous design[1]. |

illustrates in Figure 3(c), and it will coincide with the left half curve of the noise distribution in the noise rate scan approach, as shown in Figure 4(b). The threshold scan results of the noise rate scan have symmetric distribution, and the maximal noise rate, usually unpredictable in a measurement, denotes the equivalent baseline. The threshold scan should be conducted linearly and thus time-consuming. The sample accumulation also finds the equivalent baseline by locating

---
[1] This feature will be explained in Section 4.1.

– 5 –

around half of the maximal accumulated numbers. The blue curve in Figure 4(b) and the red curve in Figure 4(c) have the same transition region, which is defined as the noise width. The noise width assists the users in applying a relative threshold on the measured equivalent baseline to set the actual discriminator threshold.

The sample accumulation approach is chosen for ETROC in-pixel automatic threshold calibration because a fast scan with binary Successive Approximation Register (SAR) algorithm is possible. The total scan time can be significantly reduced with limited threshold scan points.

## 4. Sample accumulation design and FPGA verification

### 4.1 Design of sample accumulator

The core in the sample accumulation approach is the sample accumulator, as shown in Figure 5. It serves to measure the average discriminator output in a time window defined by a counter. A complete threshold calibration process requires several measurements of the average discriminator output at different discriminator thresholds to find the equivalent baseline. In Figure 5(a), the discriminator pulse is asynchronous with the 40 MHz clock. Metastability could occur if the discriminator flips at the vicinity of rising edges of the clock. A two-stage synchronizer is used to minimize the probability of counting errors due to metastability [18]. The required accumulator number of bits is one bit larger than the counter number of bits to avoid overflow. The accumulated number is stored in the "Acc" register. When the applied threshold is larger than the equivalent baseline, the discriminator outputs low, and the measured average is 0. When the applied threshold is smaller than the equivalent baseline, the discriminator outputs high, and the measured average is the maximum number defined by the length of the time window. If the measured average falls between 0 and the maximal, the applied threshold must be close or on the equivalent baseline.

An example of the sample accumulation process is presented in the timing diagram in Figure 5(b). The applied threshold is on the equivalent baseline in this example. The 40 MHz clock index represents the time window of 32768 clock periods, and the discriminator pulse is sampled at the rising edge of the 40 MHz clock. The accumulator number is reset when the sample accumulation starts and updated two clock periods later. The discriminator pulse sample is "1" at the first clock rising edge, and thus the accumulator number becomes 1. Since the second sample is "0", the accumulator number is still 1, and so on. The final accumulator number is 16381, close to half of the maximal accumulating numbers, and meets the expectations.

### 4.2 FPGA verification

The logic design of the sample accumulator has been verified in an FPGA-based test setup together with the ETROC0 chips. The test setup is shown in Figure 6. We developed two test boards for ETROC0 [13]. One is the sensor board, where an LGAD is connected to the ASIC with bonding wires. The other board is used for the ASIC alone tests, where a capacitor (2 pF) is soldered to emulate the LGAD and parasitic capacitance on the board. On the sensor board, a source meter provides a reverse high voltage (HV) to the LGAD and measures the leakage current simultaneously. A Low-voltage differential signaling (LVDS) comparator evaluation board (Part No. ADCMP604) converts the single-ended discriminator signal to a differential signal that the Altera Stratix II GX FPGA evaluation platform can take at high speed. A real-time oscilloscope (Part No. RIGOL DS1202) is used to capture the discriminator pulse for monitoring purposes.

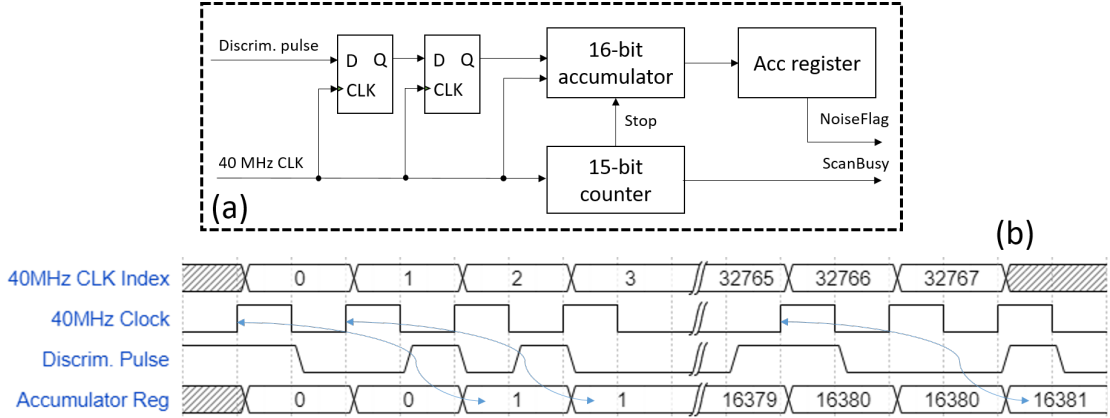

**Figure 5.** The block diagram (a) and timing diagram (b) of the sample accumulator.

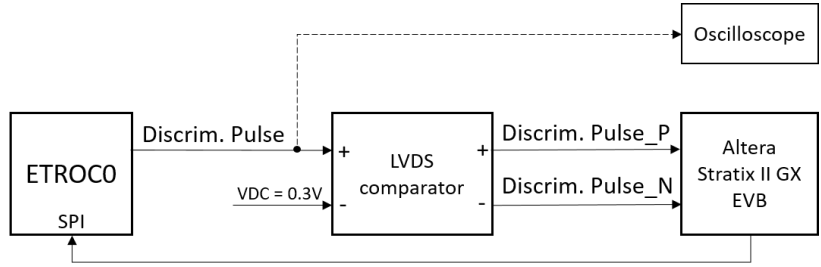

**Figure 6.** FPGA-based test setup of equivalent baseline scan.

The FPGA sets the 10-bit DAC via the Serial Peripheral Interface (SPI) in ETROC0 and handles the discriminator pulses. The output signals of the design, including the accumulator numbers, the measured equivalent baseline, and the noise width, can be observed in real-time by using the SignalTap II logic analyzer in the Quartus II software.

Figure 7(a) plots the threshold scan results versus the accumulator number of bits on the ETROC0 sensor board. The HV to the applied FBK LGAD [19, 20] is fixed at ‒330 V for a modest gain and stable sensor capacitance. The preamplifier is configured with the low-power mode (IBSel = 7) and the default feedback resistance (RFSel = 2). Figure 7(b) illustrates the measured equivalent baseline and the noise width corresponding to the curves in Figure 7(a). The equivalent baseline variation is within 1 DAC LSB (Least significant bit). The noise width increases with the accumulator bit, as expected.

The scan window length and the scan time increase exponentially with the accumulator number of bits. For the trade-off between the measured noise width and the scan window length, the accumulator number of bits in the sample accumulator is chosen as 16. In that case, the time of the single scan window is given by:

$$T_{window} = 2^{(16-1)} \times \frac{1}{40MHz} \approx 0.82 \ ms. \tag{4.1}$$

Based on this information, the total time of the in-pixel threshold calibration process will be discussed in Section 5.1.

The preamplifier baseline and its noise distribution vary with different settings, such as the bias current and the feedback resistance. The variations can also be observed with the sample accumulation algorithm. The equivalent baseline scan results versus the power modes (bias settings) on the ETROC0 sensor board are shown in Figure 8(a). The smallest IBSel code represents the largest preamplifier bias current. The measured equivalent baseline increases with the bias current, as expected. The noise width is stable regardless of the bias settings.

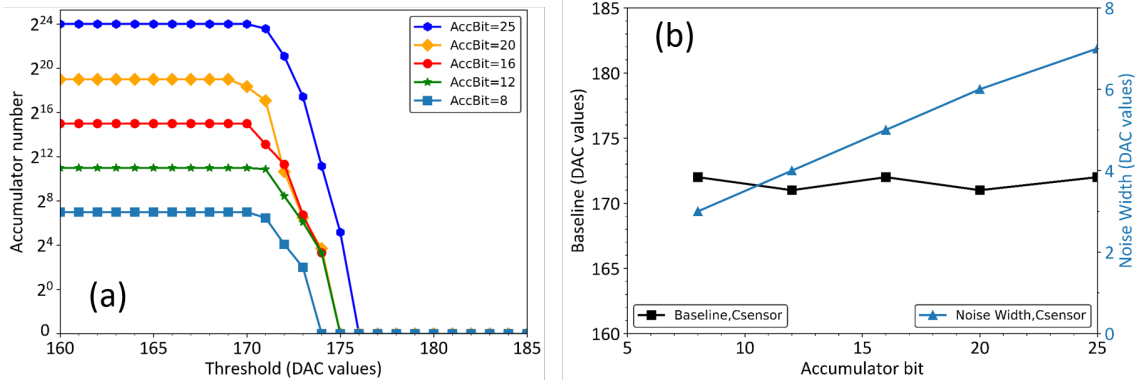

**Figure 7.** Measured results versus the threshold scanning for different accumulator bits (a); measured equivalent baseline and noise width versus accumulator bits (b) on ETROC0 with LGAD sensor capacitance.

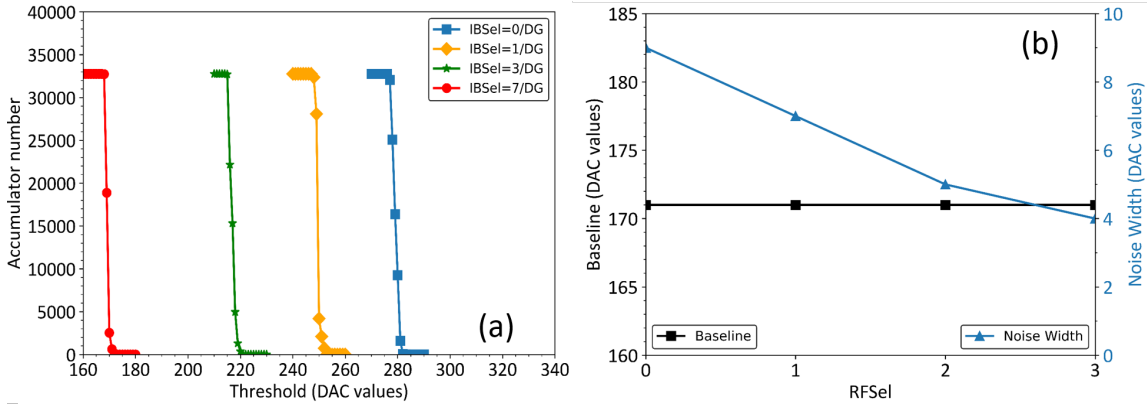

**Figure 8.** Measured equivalent baseline and noise width versus preamplifier bias settings (a), and versus preamplifier feedback resistances (b) on ETROC0 with LGAD sensor capacitance.

Figure 8(b) presents the equivalent baseline scan results versus the preamplifier feedback resistance from the sensor board. The smallest RFSel code indicates the largest feedback resistance, the largest gain in other words. No baseline shift with the preamplifier gain setting is observed, and the noise width increases with the feedback resistance, as expected.

Similar tests have been conducted on the ETROC0 board with a 2 pF soldered capacitor. The equivalent baseline for the ASIC alone is about 30 DAC LSB higher than that on the LGAD sensor, translating to a shift of 12 mV. Note that the step size of the DAC is 0.4 mV [6]. Provided that the default feedback resistance is 5.5 kΩ, an equivalent LGAD leakage current of 2.2 μA contributes to the preamplifier baseline. The measured noise width is the same on both test boards.

The comparator in ETROC discriminator has a programmable hysteresis ranging from 0 (by default) to 1 mV [13]. Figure 9 illustrates the equivalent baseline scan on the ETROC0 board with a soldered preamplifier input capacitor. The test results indicate that a higher hysteresis of the discriminator does not change the baseline but only narrows the noise width. Additionally, the measured S-curves are producible, regardless of the linear scan. The features also have been proved on the ETROC0 sensor board.

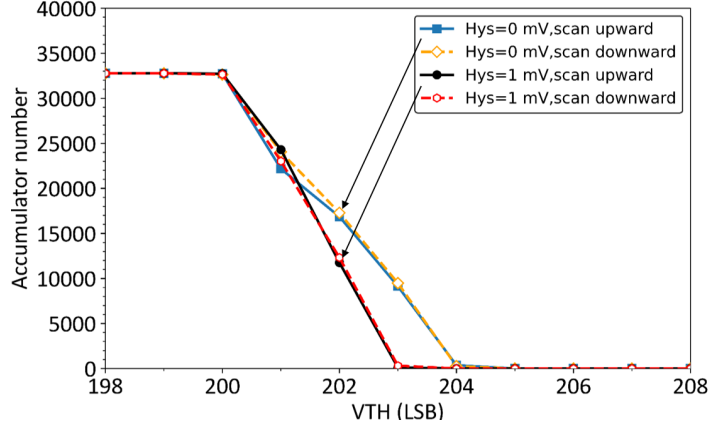

**Figure 9.** Measured results versus the threshold scanning for different discriminator hysteresis and both scan directions on ETROC0 with a 2 pF soldered capacitor.

## 5. Overall design and verification

This section presents the logic design, the implementation, and the verification of the in-pixel automatic threshold calibration.

### 5.1 Logic design

The proposed in-pixel automatic threshold calibration for ETROC instantiates the sample accumulator discussed above. The block diagram in Figure 10 presents the circuit block and its connections with a scalable I$^2$C-based slow control and the analog blocks in ETROC. The slow control block embeds the configuration and status registers in each of the 256 pixels, allowing parallel actions with a broadcast command. A rising edge of the command signal (ScanStart) launches the automatic threshold calibration procedure controlled by a state machine. The average discriminator outputs are measured at different discriminator thresholds (TH[9:0]) set by the state machine. Once the calibration is finished, an indicator signal (ScanDone) becomes high, and the equivalent baseline (BL[9:0]) and the noise width (NW[3:0]) are found and restored in I$^2$C registers. The reserved noise width number of bits is 4, based on the FPGA-based test results discussed in Section 4.2. Finally, a threshold is applied as:

$$TH[9:0] = BL[9:0] + TH\_offset[5:0], \quad (5.1)$$

where TH_offset[5:0] is a user-defined relative threshold above the measured equivalent baseline. The input 40 MHz clock is gated and only enabled during the threshold calibration for power-saving and minimizing power supply noise consideration. The calibration circuit has two reset modes: a power-on reset (POR) and the external reset via an input signal. Once reset, the scan suspends with all zero outputs.

The in-pixel automatic threshold calibration block can be bypassed, and a user-defined DAC[9:0] from the I$^2$C can be assigned to the DAC circuit. With a rising edge of ScanStart, a single average discriminator output measurement is activated in the bypass mode, and the accumulating number (Acc) is accessible through I$^2$C. The bypass function allows users to manually set the discriminator threshold and debug the possible issues during the automatic scan.

A complete threshold calibration includes a binary successive approximation scan for a rough equivalent baseline searching and a linear scan around the rough equivalent baseline. Since the 10-bit DAC sets the discriminator threshold, it takes 10 steps successive approximation to find the rough equivalent baseline, starting at the middle (TH=512) of the threshold range. After each step of discriminator pulse sample accumulation, if the accumulating number "Acc" is below half of the maximal accumulating numbers (N=32768), the next threshold will be at the middle (TH=256) of the lower half threshold range. Afterward, a rough equivalent baseline is found and stored as "BL_int" at the last step of the successive approximation scan. Since the scan directions lead to the same results proven by the FPGA test, the linear scan will be launched upward, setting the threshold from BL_int minus twelve to BL_int plus twelve. The scan range is wide enough to cover the noise width. After 25 steps of the linear scan, the equivalent baseline and the noise width are found and accessible for users to check. The threshold is updated every 1 ms in the state machine to cover a single-step scan (see eq. (4.1)) and ensure DAC output is stable before launching the measurement. It takes 35 ms to complete the threshold calibration process.

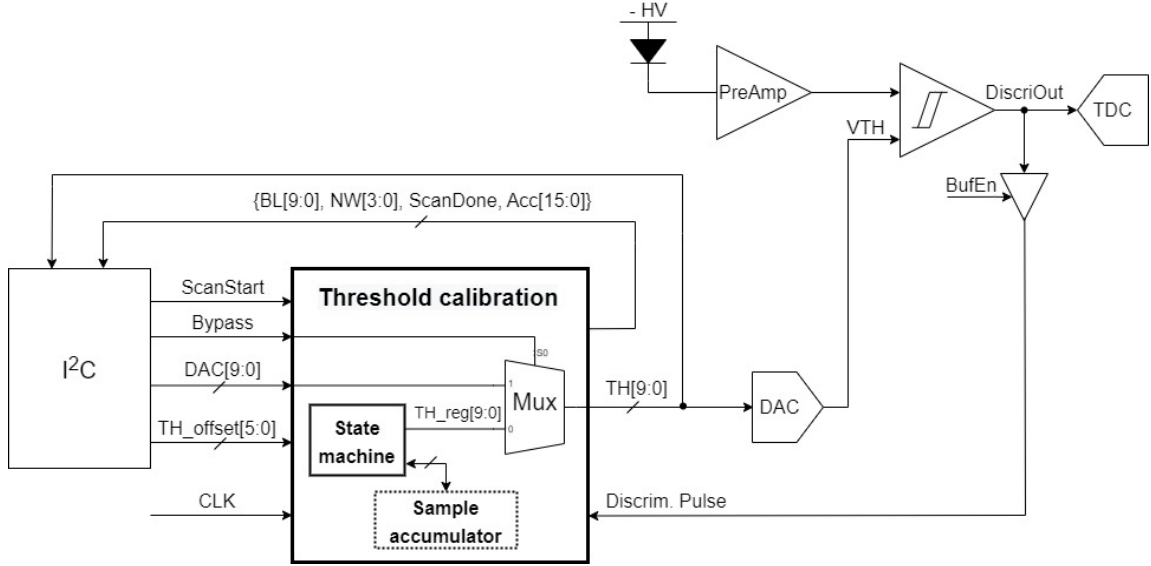

**Figure 10.** Block Diagram of in-pixel automatic threshold calibration.

## 5.2 Implementation and verification

The threshold calibration circuit has been implemented in a 65 nm CMOS process with a semi-custom design flow. Triple Modular Redundancy Generator (TMRG) tool is used for triplication and Single Event Effects (SEEs) generation [21]. Figure 11 illustrates the layout of the threshold calibration circuit. The layout area is 0.20 mm × 0.19 mm. Three full-custom POR blocks (blue delimited area) are used to generate the reset pulses at power-on. SEE of the reset could be alleviated with this triplicated design. In the worst case, the dynamic and static power consumption are estimated to be 300 μW and 10.4 μW, respectively.

A model for discriminator pulse generator is used in the test bench to improve the feasibility and efficiency of the verification. The model includes the noise distribution extracted from the real analog front-end circuits, and the equivalent baseline can be specified with DAC values. In the simulation, the equivalent baseline scan step is set to 0.25 LSB, which can be translated to 0.1 mV.

Post-layout simulations are performed with Xcelium to verify the design, where the delays due to parasitic R and C are extracted and annotated to the gate-level netlist. Figure 12 presents the post-layout simulation results of a continuous equivalent baseline scan. For the expected threshold range from 100 to 500 (DAC values), the found equivalent baseline versus the input equivalent baseline is shown in Figure 12 (a). Figure 12(b) illustrates the scanned error, which is a difference between the model baseline and the found baseline. The scanned error is within 1 LSB, indicating that the proposed automatic threshold calibration approach is effective and well adapted to the baseline variations. As an ideal model for discriminator pulse generator is used in the simulation, the results do not include the non-linearity of the DAC mainly introduced by the mismatch. In ETROC, the DAC non-linearity does not introduce calibration error as the monotonicity of the DAC is guaranteed by design [6]. The gate-level simulation with SEEs injection is conducted as well, and the majority voting and the error auto-correction are verified.

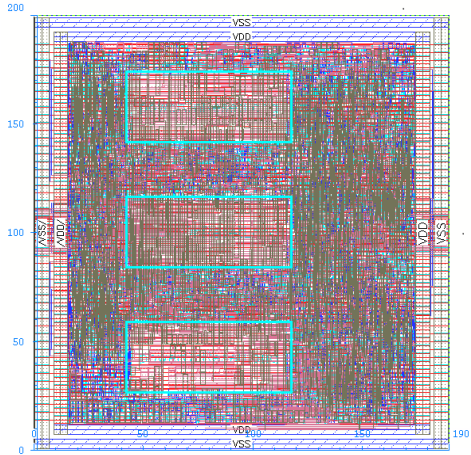

**Figure 11.** Layout of the in-pixel threshold calibration block.

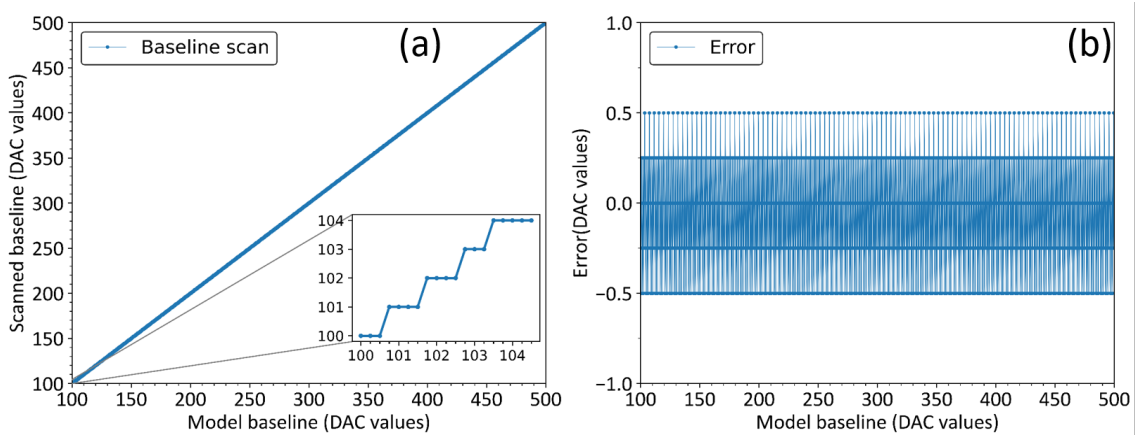

**Figure 12.** The scanned equivalent baseline (a) and the scanned error (b) versus the model equivalent baseline.

## 6. Conclusion and outlook

This paper presents the design and implementation of an in-pixel automatic threshold calibration for ETROC, a pixelated LGAD readout chip for CMS ETL HL-LHC upgrade. The significance of a proper threshold for the ETROC time measurement is elaborated. Three possible approaches for the in-pixel threshold calibration are discussed. The sample accumulation for baseline measurement is chosen to save calibration time and simplicity in implementation. Together with the ETROC0, FPGA-based tests verify the proposed approach and determine the scan window length (16-bit accumulator) and the noise width capacity (4-bit register). The threshold calibration circuit is implemented with triplication in a 65 nm CMOS process and verified in simulation. In the worst case, the dynamic and static power consumption are estimated to be 300 µW and 10.4 µW, respectively. A complete calibration process takes about 35 ms to find the equivalent baseline and apply an optimal threshold. The quantization noise dominates the found equivalent baseline error.

ETROC2, the next iteration of the ETROC series, will contain the full functionality by using the full 16×16 pixel matrix. The proposed threshold calibration circuit will be implemented inside each pixel in ETROC2. Its power supply will be in the digital domain, separated from the analog domain to avoid noise-coupling. In addition, the S-curve measurement approach can still be performed without extra circuits because the ETROC TDC can mark the valid charge injection events in a measurement window. Over an I$^2$C-based slow control at a slower speed, the S-curve measurement approach can be used as a cross-check for the threshold calibration.

## Acknowledgments

This work has been authored by Fermi Research Alliance, LLC under Contract No. DE-AC02-07CH11359 with the US Department of Energy, Office of Science, Office of High Energy Physics.

## References

[1] CMS collaboration, *A MIP Timing Detector for the CMS Phase-2 Upgrade*, CMS-TDR-020 (2019) [CERN-LHCC-2019-003].

[2] S. White, *R&D for a dedicated fast timing layer in the cms endcap upgrade*, *Acta Phys. Pol. B Proc. Suppl.* **7** (2014) 743.

[3] G. Pellegrini et al., *Technology developments and first measurements of low gain avalanche detectors (LGAD) for high energy physics applications*, *Nucl. Instrum. Meth. A* **765** (2014) 12.

[4] N. Cartiglia et al., *Design optimization of ultra-fast silicon detectors, Nucl. Instrum. Meth. A* **796** (2015) 141.

[5] W. Zhang et al., *A Low-Power Time-to-Digital Converter for the CMS Endcap Timing Layer (ETL) Upgrade, IEEE Trans. Nucl. Sci. (2021)*, early access [arXiv: 2011.01222].

[6] L. Zhang et al., *The threshold voltage generation for CMS ETL readout ASIC*, presented at 22$^{nd}$ IEEE Real Time Conference, October 12, online (2020).

[7] Y. Ding et al., *The study of calibration for the hybrid pixel detector with single photon counting in HEPS-BPIX, IEEE Trans. Nucl. Sci. (2021)*, early access [arXiv: 2011.01342].

[8] CMS Collaboration, *Commissioning and Performance of the CMS Pixel Tracker with Cosmic Ray Muons*, 2010 *JINST* **5** T03007.

[9] M. Stückelberger, *Threshold Calibration of the CMS Pixel Detector*, Diploma Thesis (2009), available at https://people.phys.ethz.ch/~ursl/home/v0/projects/stueckelberger.pdf.

[10] P. Behara, et al., *ATLAS Note: Threshold Tuning of the ATLAS Pixel Detector*, ATL-INDET-PUB-2010-001 (2010).

[11] G. Barbier, et al., *ATLAS Note: Electrical results of double-sided silicon strip modules for the ATLAS Upgrade Strip Tracker*, ATL-UPGRADE-PUB-2012-002 (2012).